\documentstyle[twoside,fleqn,espcrc2,graphicx]{article}

\newcommand{\beq}{\begin{equation}}
\newcommand{\eeq}{\end{equation}}
\newcommand{\bea}{\begin{eqnarray}}
\newcommand{\eea}{\end{eqnarray}}
\newcommand{\half}{{1\over 2}}
\newcommand{\CY}{{\cal Y} }

\newcommand{\AmS}{{\protect\the\textfont2
  A\kern-.1667em\lower.5ex\hbox{M}\kern-.125emS}}

\title{An Extended Isgur-Paton Model:  Agreement With the Lattice?}

\author{R. W. Johnson\thanks{Talk presented at Lattice 97 by R. W.
Johnson.  Sponsored by the Rhodes Trust.}
        and 
        M. Teper\address{Department of Physics, University of Oxford,
Theoretical Physics, \\
1 Keble Road, Oxford OX1 3NP, United Kingdom}}
       
\begin{document}

\thispagestyle{empty}

\begin{abstract}
The spectrum for the pure gauge sector is calculated for an extended
Isgur-Paton model in 2+1 and 3+1 dimensions and compared to recent lattice
calculations of the glueball spectrum.  The IP model is extended by
inclusion of a rigidity (curvature) term and, in D=2+1, mixing through a higer
topological contribution.  For a choice of parameterizations, near
quantitative agreement is found for SU(3) in D=2+1, but in D=3+1 the
extensions fail to remedy the qualitative disagreement.
\end{abstract}

\maketitle

\section{Introduction}

The traditional Isgur-Paton model \cite{IP} was inspired from the
strong-coupling limit of lattice QCD.  It describes a pure glue state as a
closed loop of chromoelectric flux of vanishing diameter.  The model is
nonrelativistic and invokes an adiabatic limit such that motion is split
between "fast" transverse oscillcations of the string and "slow" radial
excitations.  For $SU(N\geq3)$ the string carries an orientation vector.
The predictions of the model depend greatly on the number of spatial
dimensions, and so we will discuss the D=2+1 case primarily.  Adiabatically 
separating $\rho_0$ allows quantization of the vibrational modes into phonons
characterized by quantum occupation number $n_m^\pm$, where $\pm$ indicates 
the "helicity"  of the mode.  We also should expect a contribution from
the rest energy of the string, proportional to the string tension
$\sigma$.  For a loop of radius $\rho_0$, we have the Hamiltonian
\beq H(\rho_0) = 2\pi\sigma\rho_0 + \frac{\sum_{i,m}\pi_{i,m}^2}
{2\pi\sigma\rho_0} + \frac{\pi\sigma}{2\rho_0} \sum_{i,m} m^2 q_{i,m}^2,
\label{eq:ham1} \eeq
where $q_{i,m}$ represents the normal coordinates for $\alpha_m$ and
$\beta_m$ and $\pi_{i,m}$ is the conjugate momentum.  The zero-mode energy
diverges but may be regularized by imposing a short-scale cutoff $a$ on
the string.  Such a discretization leads to a zero-mode contribution as
$\frac{8\rho_0}{a^2} - \frac{13}{12\rho_0} + \cdots.$  The term linear in 
$\rho_0$ can be identified with a "renormalization" of
the string tension $\sigma$.  The term linear in $1/\rho_0$ has been
identified as a L\"{u}scher correction term, but inclusion of a rigidity
term before quantization provides a $1/\rho_0$ dependence, and thus this
term can be more easily identified with a renormalization of the effective
elasticity in the extended model.  The phonon modes can be shown to contribute 
to the energy as $m(n_m^+ + n_m^-)/\rho_0$, giving the mode energy factor 
$M=\sum_m m(n_m^+ + n_m^-)$.  For D=2+1, all angular momentum arises from the 
phonons, thus we have $J=\Lambda=\sum_m m(n_m^+ - n_m^-)$.

Since we expect the picture of the flux tube to break down near the
origin, as well as to soften the singularity in the potential at
$\rho_0=0$, the potential in our effective Hamiltonian will be multiplied
by a heuristic suppression factor, $F(\rho)$.  Originally in \cite{IP}, 
$F(\rho)=(1-e^{-f\sqrt{\sigma}\rho})$, where $f$ is a free parameter near unity, but 
we find that a rational function, $F(\rho)=(1-f/(\rho+f))$, gives a better
fit to the lattice data.

\section{Inclusion of Elasticity}

Both intuition and calculation conclude that a flux tube should have
nonvanishing thickness \cite{WG} in the transverse dimension.  One manifestation of finite thickness
we may expect to find even in our nonrelativistic model is an effective
elasticity of the flux tube.  The free energy per unit length of a rigid bent rod is $\half \CY
\frac{1}{r^2} I$, where $\CY$ is the Young's modulus and $I$ is the
moment of intertia perpendicular to the axis of bending.  We expect
$\CY$ and $I$ to be constant along the flux tube, thus integrating
along the length of the flux tube
$\oint \half \CY I \frac{1}{\rho_0^2} dl \propto {1\over \rho_0} \:.$  
In string language this is the curvature term.  The constant of 
proportionality is our effective elasticity, denoted
$\gamma$, and will be a free parameter in our action.  We can see now
that the effect of discretization on a scale $a$ is to renormalize the
elasticity 
$\gamma \rightarrow \gamma - \frac{13}{12} \: . $
There has been some discussion \cite{KC} as to the sign such an elasticity
should take.  We leave the elasticity as a free parameter letting the
best fit parameter indentify the correct sign.

\section{Mixing}

Under charge conjugation, $C$, the loop orientation flips.  To split the $C=+$ 
and $C=-$ sectors, we must introduce mixing between "left" and "right" 
orientated loops.  We explore two scenarios:  direct mixing and mixing through a 
higher topological state.  Writing the independent orientations of the glue 
loop as $\psi_L$ and $\psi_R$, the direct mixing matrix Hamiltonian is 
given in equation (\ref{eq:HM1}). For 
$\Psi = \left( \begin{array}{lr} \psi_L & \psi_R \end{array} \right)^T$,
\beq \Psi^\dag H_1 \Psi = 
\Psi^\dag
\left[ \begin{array}{cc}
       H_{L,R}    &  -\alpha \\
       -\alpha    &  H_{L,R}
       \end{array} \right]
\Psi , \label{eq:HM1} \eeq
where $\alpha$ is the mixing parameter and $H_{L,R}$ is the Hamiltonian
for a single glue loop.  Since we would expect mixing to occur more easily
for smaller loops, $\alpha$ may be multiplied by the complement of the 
suppression function of the potential, so that the mixing goes as 
$\alpha(1-F(\rho))$.  Eigenstates for $C$ are built as $\psi_L + \psi_R$ and
$\psi_L - \psi_R$. 

We may instead include a state, denoted $\psi_Y$, consisting of a glue
loop with an extra diameter as an intermediate state through which one
orientation may fluctuate to the other.  Our matrix Hamiltonian then
reads, for 
$\Psi = \left( \begin{array}{lcr} \psi_L + \psi_R & \psi_L -
\psi_R & \psi_Y \end{array} \right)^T$,
\beq 
\Psi^\dag H_2 \Psi =
\Psi^\dag
\left[ \begin{array}{ccc}
H_{L,R}   &   0           &   \alpha \\
0         & H_{L,R}       &   \eta_C \alpha \\
\alpha    & \eta_C \alpha & H_Y \end{array} \right]
\Psi , \label{eq:HM2} \eeq
where the factor $\eta_C = \pm 1$ arises from coupling the $\psi_Y$ state
to the appropriate admixture of $\psi_L$ and $\psi_R$ to match the C
eigenvalue.  Since the $\psi_Y$ state is not rotationally invariant, we
must write the linear superposition as
$\psi_Y = \int d\theta e^{iJ\theta} |\psi_{Y \theta=0} \rangle$.  A quick
calculation shows that the $\psi_Y$ state has C eigenvalue $C_Y = (-1)^J$. 
The combinations $\psi_L \pm \psi_R$ have opposite C eigenvalues, thus by
including the factor $\eta_C = (-1)^J$ the $\psi_Y$ state decouples from
the admixture with the wrong C eigenvalue.  As above, $\alpha$ may multiply 
a function of $\rho$, but the energetic cost of creating the diameter for 
large $\rho_0$ provides an inherent suppression of mixing for large $\rho_0$.

\section{Building the Hamiltonian}

We build our Hamiltonian from (\ref{eq:ham1}) invoking an adiabatic
separation of the variable $\rho$, as in the case of mesons \cite{IP}.
When we take ${\pi_{\rho_0}}^2 \rightarrow - \frac{d^2}{d \rho_0^2}$, we 
identify our kinetic terms as
\bea H^{L,R}_{\mathit{kinetic}} & = & \frac{-1}{2(2 \pi \sigma \rho_0)}
\frac{d^2}{d \rho_0^2} \\
H^Y_{\mathit{kinetic}}        & = & \frac{-1}{2(2 \pi \sigma \rho_0 +  2v)}
\frac{d^2}{d \rho_0^2}  .
\eea
The potential $V(\rho_0)$, given by the rest energy of the string and the
contributions from the rigidity term and phonon modes, is 
\bea V^{L,R} & = & 2 \pi \sigma \rho_0 + \frac{M + \gamma}{\rho_0} \\
     V^{Y}   & = & (2\pi + 2) \sigma \rho_0 + 2v + \frac{\gamma}{\rho_0} .
\eea
The term $2v$ in $V^Y$ is a contribution from the vertices each of static 
energy $v$.  Such a contribution may follow from considerations in \cite{MR}.  
We also must include the rotational contribution for $\psi_Y$,
\beq H^Y_{\mathit{rotational}} = \frac{L^2}{2I_Y} = \frac{J^2}{(4\pi +
{4\over 3}) \sigma \rho_0^3 + 2v \rho_0^2}, \eeq
but we do not include any phonon contributions. 

\section{Results}

\begin{figure}[t]
\centering
\includegraphics[width=75mm,trim=20 68 0 0]{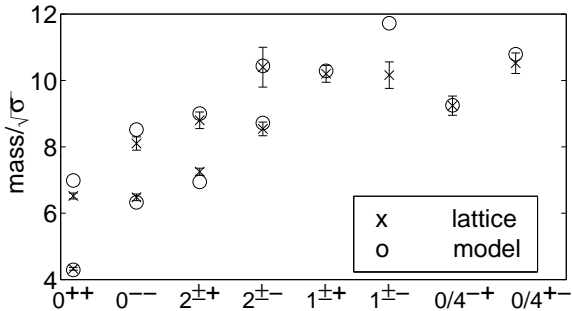}
\caption{Spectrum for direct mixing mechanism.  $f=0.17$, $\gamma=0.90$, 
and $\alpha=4.5$. $\chi^2 / \mbox{dof} = 44.1/(12 - 3) = 4.9$.}
\label{fig:allold}
\end{figure}

The spectrum is calculated by a suitable numerical diagonalization of the 
Hamiltonian and compared to recent lattice results (see, for example, 
\cite{MT}).  We minimize $\chi^2$ with respect to the $J=0,2$ sectors by a 
simplex search in parameter space and then calculate the full spectrum. Some 
ambiguity arises when identifying certain $J^{PC}$ states for comparison.
Lattice Monte Carlo simulation gives a $0^{-+}$ below the $1^{++}$, whereas 
our model predicts the lightest $0^{-+}$ (at $M=8$) well above the $1^{++}$.
However, our model does predict a $4^{-+}$ below the $1^{++}$, suggesting 
that the Monte Carlo signal may be a $J=4$ in disguise, a possibility for an 
Euclidean square lattice.  Figure~\ref{fig:allold} summarises our results 
for the direct mixing mechanism, with $\alpha$ multiplied by $f/(\rho + f)$.  
Except for the $1^{\pm-}$, the model and the lattice are in remarkable 
agreement, although $\chi^2$ is somewhat larger than a believable fit should 
have.  The degeneracy of the lattice data for the $J=1$ sector is not 
available to the model, which maintains a significant splitting for all $J$ 
such that the $C=+$ states are lighter than the corresponding $C=-$.

\begin{figure}[t]
\centering
\includegraphics[width=75mm,trim=20 68 0 0]{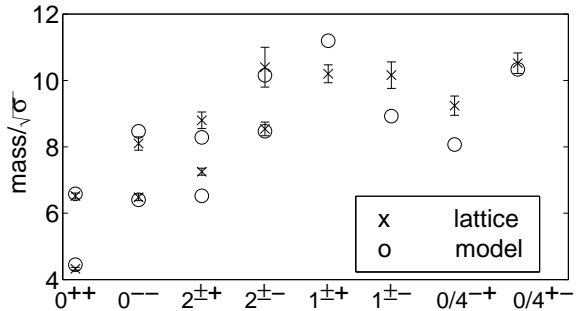}
\caption{Spectrum for Y state mixing mechanism.  $f=0.28$, $\gamma=2.4$, 
$\alpha=3.8$, and $v=9.3$.  $\chi^2 / \mbox{dof} = 64.7/(12 - 4) = 8.1$.}
\label{fig:allnew}
\end{figure}

For mixing through $\psi_Y$, our results are shown in Figure~\ref{fig:allnew}, 
with no multiplicative suppression of $\alpha$.  Generally we find a poorer fit 
despite the additional parameter.  Here we notice an interesting
feature of this mechanism, namely that for even $J$ the $C=+$ states are 
lighter, but for odd $J$ the $C=-$ states are lighter.  For the $J=1$ sector
the lattice data must be averaged across parity to make our comparison, and
the $1^{\pm-}$ might move relative to the $1^{\pm+}$ as better data become 
available.  Which direction the $1^{\pm-}$ moves (if at all) would provide 
a reasonable criteria for distinguishing between mixing mechanisms.

\section{Conclusions}

The Isgur-Paton model, extended by inclusion 
of an effective elasticity and by various mixing mechanisms, agrees reasonably
with the lattice SU(3) spectrum in D=2+1, suggesting that a flux tube picture 
for glueballs might be viable there.  In D=3+1, a qualitative discrepancy 
remains, as the model predicts a low-lying 
$1^{-+}$ as an orbital excitation of the ground state $0^{++}$ which is at odds
with the relatively heavy $1^{-+}$ on the lattice.  Nonetheless, the quality 
of the comparison in D=2+1 suggests futher work be done mapping the pure 
gauge spectrum for this dimensionality.  An interesting point for future 
comparison would be the mass of the $J=3$ sector, which the model predicts to 
be signficantly lighter than the $J=1$ sector.

We thank J. Paton for many useful discussions.

\end{document}